\documentclass[superscriptaddress,aps,pra,twocolumn,showpacs,nofootinbib,longbibliography]{revtex4-1} 
\usepackage{amsmath}
\usepackage{latexsym}
\usepackage{amssymb}
\usepackage{amsthm} 
\usepackage{mathrsfs}
\newtheoremstyle{mystyle}{}{}{}{}{\bfseries}{. }{0pt}{\thmname{#1}\thmnumber{ #2}\thmnote{ (#3)}}
\theoremstyle{mystyle}

\usepackage[colorlinks=true,citecolor=blue,urlcolor=blue]{hyperref}
\usepackage{color}
\usepackage{graphics,epstopdf}
\usepackage{soul}
\usepackage[demo]{graphicx}
\usepackage{caption}
\usepackage{lipsum,mwe}
\usepackage{subcaption}
\usepackage{lipsum}
\usepackage{ragged2e}

\begin{document}

\title{Unbounded-input explicit Bell inequalities for general quantum networks}

\author{Yao Xiao}
\affiliation{School of Mathematical Sciences, Beijing University of Posts and Telecommunications, Beijing 100876, China}
\affiliation{Key Laboratory of Mathematics and Information Networks, Beijing University of Posts and Telecommunications, Ministry of Education, China
}
\affiliation{State Key Laboratory of Networking and Switching Technology, Beijing University of Posts and Telecommunications, Beijing 100876, China}
\author{Fenzhuo Guo}
\email{gfenzhuo@bupt.edu.cn}
\affiliation{School of Mathematical Sciences, Beijing University of Posts and Telecommunications, Beijing 100876, China}
\affiliation{Key Laboratory of Mathematics and Information Networks, Beijing University of Posts and Telecommunications, Ministry of Education, China
}
\affiliation{State Key Laboratory of Networking and Switching Technology, Beijing University of Posts and Telecommunications, Beijing 100876, China}
 \author{Haifeng Dong}
 \affiliation{School of Instrumentation Science and Opto-electronics Engineering, Beihang   University, Beijing
 	100191, China}
 \author{Fei Gao}
 \affiliation{State Key Laboratory of Networking and Switching Technology, Beijing University of Posts and Telecommunications, Beijing 100876, China} 

\begin{abstract}
Quantum nonlocality in networks featuring multiple independent sources underpins large-scale quantum communication and poses fundamental challenges for its characterization. 
In this work, we construct a family of explicit nonlinear Bell  inequalities to verify the nonlocality across the general multi-input quantum networks. The construction of these inequalities relies on the number of leaf nodes, a network parameter that can be identified by a linear-time algorithm. Our approach establishes a structural connection between bipartite full-correlation Bell inequalities and network Bell inequalities, enabling the analytical derivation of optimal quantum violations and the conditions under which they occur. We further quantify the upper bound on maximal violations achievable by arbitrary two-qubit mixed states in such networks, under separable measurements, and evaluate the noise robustness of the proposed inequalities via the visibilities of Werner states.
Finally, we demonstrate that these inequalities can, in a device-independent manner, distinguish between network topologies of equal size that differ in the number of leaf nodes. 
\end{abstract}

\maketitle

\section{Introduction} \label{1}
Spatially separated observers performing local measurements on a shared entangled quantum state can establish strong correlations that cannot be explained by
any local hidden variable theory \cite{bell64}. This  phenomenon, referred as quantum nonlocality  \cite{review}, lies at the heart of quantum theory and can be verified through the violation of suitable Bell inequality. Beyond its fundamental significance, nonlocality serves as a vital resource for device-independent quantum tasks, such as quantum key distribution \cite{QC2007}, random number generation \cite{rg2010}, and reduction of communication complexity \cite{CC}. 

Quantum nonlocality arisen in single-source scenarios has been
extensively studied. In such scenarios, the achievable local correlations forms a convex hull \cite{review},  and numerous linear Bell inequalities have been proposed to verify the nonlocality \cite{Stand1,Stand2,review,facets}. 
Network scenarios with multiple independent sources give rise to correlations that exhibit nonconvex features, making the construction of Bell inequalities for detecting network nonlocality more challenging \cite{reviewnet2022}.  
Pioneering works by Branciard $et$ $al$. \cite{bilocPRL,bilocPRA} established nonlinear Bell inequalities to verify nonbilocal correlations in the simplest quantum network. Subsequent works generalized this simplest scenario to several typical multi-source network structures, like star-shaped \cite{star2024cut,starPRA,star2017,star2017map,star2021,star2022multi}, chain-shaped \cite{chainQIP,chain2022}, and tree-shaped networks\cite{treePRA,treemulti}.  
However, these works are specific to particular topologies, while deriving inequalities applicable to general network structures holds greater practical relevance.

Several methods have been developed to derive Bell inequalities for general networks. Refs.~\cite{Polynomial2016} and \cite{AG} employed linear programming technique and algebraic geometry methods, respectively, to construct Bell inequalities for small-scale general networks. An iterative approach was introduced in Ref.~	\cite{iterative2016}, where starting from a Bell inequality for a given network, one can construct a new one for an extended network obtained by adding a bipartite source. This iterative procedure was further shown to, in principle, allow the derivation of Bell inequalities for any acyclic networks with multipartite sources \cite{acyclic2016}. However, these construction methods do not provide explicit expressions for general networks and instead require case-by-case derivations, which limits their practical applicability. In 2018, Luo derived explicit Bell  inequalities for general $N$-party networks with two inputs per party, based on identifying the largest set of mutually unconnected parties \cite{Luo2018}. A computationally efficient algorithm with time complexity $O(N^{2.5})$ was also introduced in Ref.~\cite{Luo2018} to find this set, though it may yield suboptimal results in large-scale or structurally complex networks.

In this work, we present efficient and simpler explicit nonlinear Bell inequalities for general $N$-party networks with $M$ independent sources. The construction relies on identifying the number of leaf nodes, which can be efficiently computed in linear time $O(N+M)$. Our framework accommodates arbitrary inputs per party, under the assumption that all intermediate nodes receive the same number of inputs. We establish a direct link between the bipartite full-correlation Bell inequalities (FCBIs) associated with peripheral sources and the network inequality.  
To verify the nonlocality in noisy quantum networks using our inequalities, we derive the upper bound on maximal quantum violation for any ensemble of two-qubit mixed states produced by sources and measured with separable measurements, together with the conditions required to achieve it. 
Specifically, considering all sources emit Werner states, we determine the constraints on visibilities under which network nonlocality can be revealed.
Finally, we show that our inequalities provide an effective tool for distinguishing networks with same size but different leaf-node counts, including cases that cannot be discriminated by methods presented in Ref.~\cite{witness}.

\section{Explicit bell inequalities}
 
\begin{figure}[]
	\resizebox{6.2cm}{6.2cm}{\includegraphics{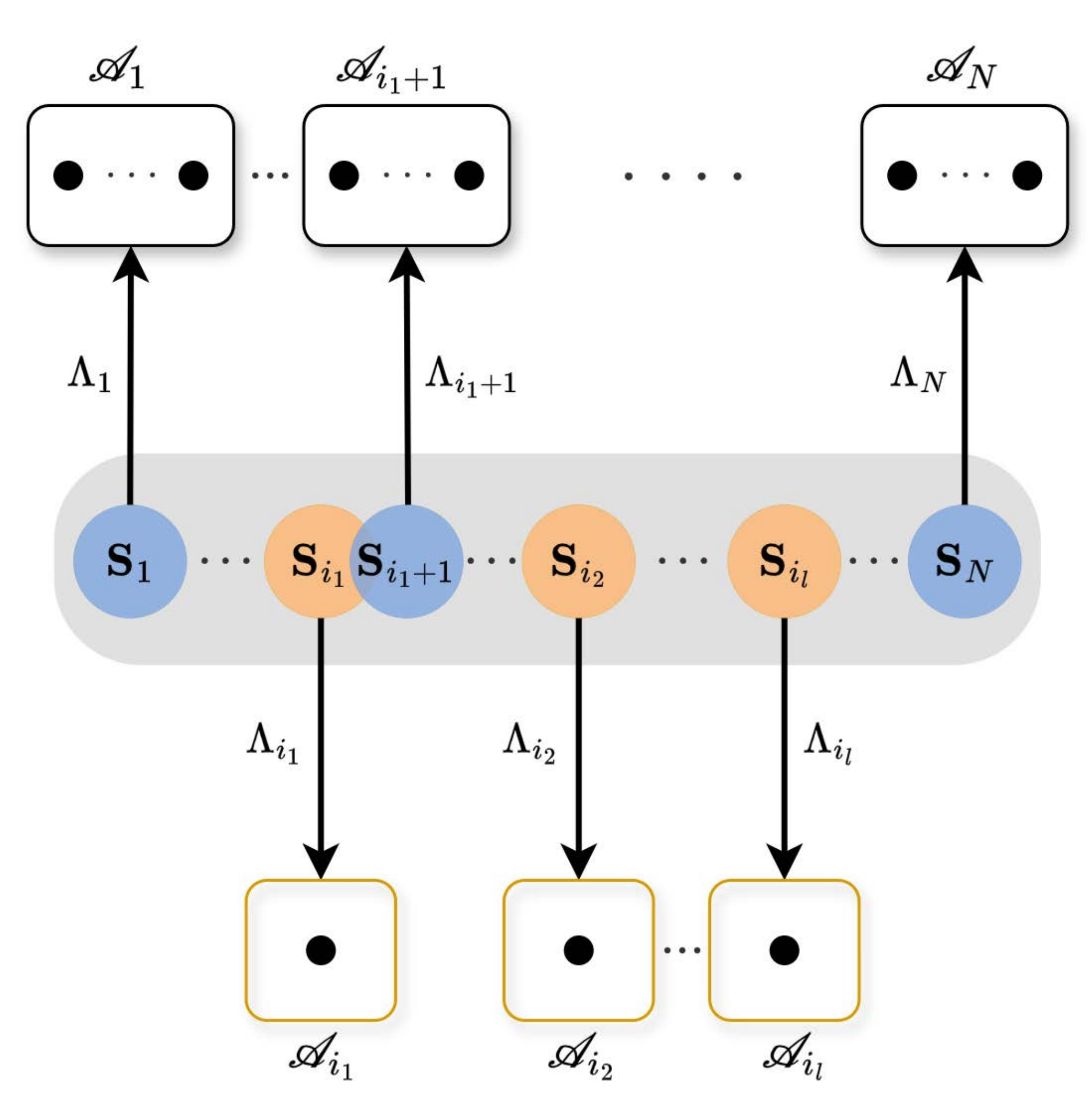}}
	\caption{\footnotesize\justifying 
	Schematic diagram of general network with $l$ leaf nodes (orange squares). There are $N$ parties (nodes) $\mathscr{A}_1$, $\mathscr{A}_2$, \ldots, $\mathscr{A}_N$ and $M$ bipartite sources $\mathscr{S}_1$, $\mathscr{S}_2$, \ldots, $\mathscr{S}_M$. Each party $\mathscr{A}_i$ can receive particles from at most $M$ sources $\mathbf{S}_i=\{\mathscr{S}_{j_1},\mathscr{S}_{j_2},\ldots,\mathscr{S}_{j_{s_i}}\}$ that distribute hidden variables $\Lambda _i=\{\lambda_{j_1}, \lambda_{j_2},\ldots, \lambda_ {j_{s_i}}\}$, where $\cup_{i=1}^{N}\mathbf{S}_i=\{\mathscr{S}_{1},\mathscr{S}_{2},\ldots,\mathscr{S}_{M}\}$. Intermediate nodes (black squares) receive multiple particles, while each leaf node  $\mathscr{A}_{i_{\ell}}$  only receives one particle from peripheral source $\mathscr{S}_{i_{\ell}}$ (orange circle), i.e., $\mathbf{S}_{i_{\ell}}=\{\mathscr{S}_{i_{\ell}}\}$, $\ell=1,2,\ldots,l$.}  
    \label{fig1}
\end{figure}

Consider a general network depicted in Fig. \ref{fig1} consisting of $N$ parties $\mathscr{A}_1$,\ldots, $\mathscr{A}_N$  which are mediated by $M$ independent bipartite sources $\mathscr{S}_1$,\ldots, $\mathscr{S}_M$. Each source connects a distinct pair of parties and together forming the overall network.  Labeling the inputs for party $\mathscr{A}_{i}$ by $x_{i}\in\mathcal{X}_{i}$ with binary outputs $a_{i}\in\{0,1\}$. 
The correlation generated in the network is local if the joint probability distribution admits the following form:
\begin{eqnarray}\label{LHV}
	P(\vec{a}|\vec{x})&=&\int \left(\prod_{j=1}^{M}d\lambda_j q_j(\lambda_{j})\right)\prod_{i=1}^{N}p(a_i|x_i,\Lambda_{i}),
\end{eqnarray} 
where $\vec{a}=(a_{1},a_2, \ldots,a_{N})$,  $\vec{x}=(x_{1},x_2, \ldots, x_{N})$, $\lambda_j$ denotes the hidden variable distributed by source $\mathscr{S}_j$ and follows the distribution $q_j(\lambda_j)$, satisfying the normalization
condition $\int d\lambda_j q_j(\lambda_j)=1$.  $\Lambda _i=\{\lambda_{j_1}, \lambda_{j_2},\ldots, \lambda_ {j_{s_i}}\}$ denotes the set of classical variables from  sources $\mathbf{S}_i=\{\mathscr{S}_{j_1},\mathscr{S}_{j_2},\ldots,\mathscr{S}_{j_{s_i}}\}$ associated with party $\mathscr{A}_{i}$. If the probability distribution cannot be decomposed into Eq.~(\ref{LHV}), it is said to be nonlocal. Any distribution that can be written in the above form must satisfy a suitable network Bell inequality, which can be violated by measuring quantum states distributed in network.

Leveraging geometric features of network topologies, we construct explicit nonlinear Bell inequalities for general quantum networks. In our framework, parties connected to a single source are referred to as leaf nodes, with the associated source termed a  peripheral source. For clarity, we label by $\mathscr{A}_{i}$ the leaf node connected to peripheral source $\mathscr{S}_{i}$, and denote the set of indices of all peripheral sources as $\mathcal{I}$.  
In a network with $N$ nodes and $l$ $(l\geqslant2)$ leaf nodes,
we denote $\Gamma=\{i_1,i_2,\ldots,i_l\}$ the set of indices of all leaf nodes, and $\overline{\Gamma}=[N]\backslash\Gamma$ the set of indices of other nodes (i.e., intermediate nodes). By definition, $\Gamma=\mathcal{I}$. Assume that the number of inputs of the intermediate nodes is the same, i.e., $\{|\mathcal{X}_i|=k,i\in\overline{\Gamma}\}$. Let $A_{x_i}$ be the measurement of the party $\mathscr{A}_i$ with input $x_i$.
Any correlation generated from a network (see Fig. \ref{fig1}) with classical variables is bounded by the following nonlinear Bell inequality, 
\begin{eqnarray}\label{S}
	\mathcal{S}=\sum_{j=1}^{k}|I_j(N,l)|^{\frac{1}{l}}\leqslant \Big(\prod_{i\in\Gamma}\beta_i\Big)^\frac{1}{l},
\end{eqnarray}
where the correlator $I_j(N,l)$ is defined as
\begin{eqnarray}\label{I}
	I_j(N,l)=\sum_{x_i,i\in\Gamma}\Big(\prod_{i\in\Gamma}M_{x_i,j}\Big)\langle A_{x_1}A_{x_2}\cdots A_{x_N}\rangle,
\end{eqnarray}
and $\langle A_{x_1}A_{x_2}\cdots A_{x_N}\rangle=\sum_{\vec{a}}(-1)^{\sum_{i=1}^{N}a_i}p(\vec{a}|\vec{x})$.
The quantity $\beta_i$ is the classical bound of bipartite FCBI $\mathcal{B}_i$  associated with peripheral source $\mathscr{S}_i$, which is fully characterized by a real coefficient matrix $\mathbf{M}^{(i)}=(M_{x_i,j})$ (See Appendix \ref{appe11} for the detail).  
Note that for different $j$, define the correlator using different inputs $x_i$ of all intermediate parties $\{\mathscr{A}_i,i\in\overline{\Gamma}\}$. Using $A_j^{u}$ to represent $A_{x_u=j}$ and denoting $\Delta_j^{i}=\sum_{x_i}M_{x_i,j}A_{x_i}$ for all $i\in\Gamma$, the term in Eq. (\ref{I}) is rewritten as $I_j(N,l)=\Big\langle\prod_{u\in\overline{\Gamma}}A_j^{u}\prod_{i\in\Gamma}\Delta_j^{i}\Big\rangle$.

Coefficients $\{M_{x_i,j}\}$ in inequality (\ref{S}) are determined from bipartite FCBIs $\{\mathcal{B}_i\}$ related to the number of measurement settings for the parties connected by peripheral sources $\{\mathscr{S}_i,i\in\Gamma\}$. For example, when both nodes connected via peripheral source $\mathscr{S}_i$ have two inputs, CHSH inequality is considered for $\mathscr{S}_i$ such that $M_{x_i,j}=\frac{1}{2}(-1)^{x_i\cdot j}$  ($x_i,j\in\{1,2\}$). Further examples are provided in Appendix~\ref{exam}.
Another key quantity $l$ corresponds to the number of degree-one vertices in the underlying network graph. It can be computed in linear time $O(N+M)$ using an adjacency list representation \cite{leaf}, as each bipartite source contributes one edge, making the number of edges equal to $M$.
    
For quantum networks with identical topology but replacing all classical variables $\lambda_1,\ldots,\lambda_M$ with bipartite quantum states $\rho_1,\ldots,\rho_M$, the quantum correlations are bounded by 
\begin{eqnarray}\label{SQ}
	\mathcal{S}=\sum_{j=1}^{k}|I_j(N,l)|^{\frac{1}{l}}\leqslant\Big(\prod_{i\in\Gamma}\big(\mathcal{B}_i\big)_Q^{\text{opt}}\Big)^\frac{1}{l},
\end{eqnarray}
where $(\cdot)_Q^{\text{opt}}$ denotes the optimal quantum value of $\cdot$. 
The network correlators can be expressed as 
$\langle\otimes_{i=1}^{N}A_{x_{i}}\rangle_{\rho} =\text{Tr}[(\otimes_{i=1}^{N}A_{x_{i}})(\otimes_{j=1}^{M}\rho_{j})]$, where $A_{x_{i}}$ is the dichotomic measurement for party $\mathscr{A}_i$. The upper bound in Eq.~(\ref{SQ}) is derived using the elegant sum-of-squares (SOS) approach introduced in \cite{star2021}, with full proof given in Appendix \ref{appe13}.
  
In what follows, we show that the upper bound can be attained using separable measurements. Consider the bipartite state $\rho_i$ in network, shared between parties $\mathscr{A}_\square$ and $\mathscr{A}_\triangle$. The local observables acting on these two subsystems are denoted by $A_{x_\square}^{(i)}$ for $\mathscr{A}_\square$ and $B_{x_\triangle}^{(i)}$ for $\mathscr{A}_\triangle$, respectively. Intermediate parties perform separable measurements on each of their subsystems. For an intermediate party $\mathscr{A}_i$$(i\in\overline{\Gamma})$, the measurement operator admits the factorized form   $A_{x_i}=\prod_{u\in U\subset\mathcal{I}_i}A^{(u)}_{x_i}\prod_{v\in \mathcal{I}_i\backslash U}B^{(v)}_{x_i}$, where $\mathcal{I}_i$ denotes the set of indices of sources associated with $\mathscr{A}_i$.  
For a leaf node $\mathscr{A}_i$$(i\in\Gamma)$, the measurement reduces to that on its single subsystem, namely  $A_{x_i}=A^{(i)}_{x_i}$.  As specified earlier, $\mathcal{I}=\Gamma$.
The upper bound is achieved under the following conditions.

(1) For each peripheral source $\mathscr{S}_i$ $(i\in\mathcal{I})$, the local observables $\{A_{x_i}^{(i)}, x_{i}\in\mathcal{X}_{i}\}$ and $\{B_{j}^{(i)}, j\in[k]\}$ acting on  $\rho_i$ (i) maximize
$\big(\mathcal{B}_{i}\big)_Q^{\text{opt}}$ and (ii) either satisfy $\text{Rank}(\mathbf{X})=1$ or, for some $i\in\mathcal{I}$, $x_{j,i}=0$ for all $j\in[k]$, where $\mathbf{X}=(x_{j,i})$ with $x_{j,i}=\sqrt{\text{Tr}[(\Delta_j^{i})^{\dagger}\Delta_j^{i}\rho_i]}$ and $\Delta_j^{i}=\sum_{x_i}M_{x_i,j}A_{x_i}^{(i)}$.  

(2) For each intermediate source $\mathscr{S}_i$ $(i\in[M]\backslash\mathcal{I})$, the state $\rho_i$ is measured by local observables $\{A_{j}^{(i)},j\in[k]\}$ and $\{B_{j}^{(i)}, j\in[k]\}$ such that 
$\langle A_{j}^{(i)}B_{j}^{(i)}\rangle_{\rho_{i}}=1$ for all $j\in[k]$.
 
Eqs.~(\ref{S}) and (\ref{SQ}) establish a direct link between bipartite FCBIs and the constructed network Bell inequality. Both the  classical bound and optimal quantum violation of (\ref{S}) can be determined using the results of well-developed bipartite FCBIs. 

For many FCBIs, such as the CHSH inequality \cite{CHSH}, the chained Bell inequality \cite{CBI}, Gisin's elegant Bell inequality (EBI) \cite{EBI}, and the Bell inequalities derived from communication games \cite{2020BIRAC}, self-testing protocols show that the quantum state and measurements yielding the maximal violation $\big(\mathcal{B}_{i}\big)_Q^{\text{opt}}$ ensure that $\sqrt{\text{Tr}[(\Delta_j^{i})^{\dagger}\Delta_j^{i}\rho_i]}$ takes the same value for all $j$ \cite{SelfEBI,CBIself,CHSHself,Self2020BIRAC}. Hence, condition (1) required to achieve the optimal quantum value $(\prod_{i\in\Gamma}\big(\mathcal{B}_i\big)_Q^{\text{opt}})^\frac{1}{l}$ is automatically satisfied. Condition (2) can be fulfilled by measuring  all systems associated with intermediate sources using fixed local observables along $\sigma_3$ and by assuming that the correlation matrix of each intermediate state has unit largest singular value. This further implies that the optimal quantum violation of inequality (\ref{S}) can be achieved even when intermediate sources emit classical states $\rho_u$ with correlation matrix $diag\{0,0,1\}$ (e.g., $\rho_u=|00\rangle\langle00|$).  
 
In contrast to previous nonlinear Bell inequalities designed for specific network topologies~\cite{star2017,star2017map,star2021,star2022multi,starPRA,chainQIP,chain2022,treePRA,treemulti}, our inequalities in Eq.~(\ref{S}) are applicable to general network
structures with multiple leaf nodes.   By specifying appropriate parameters in inequality~(\ref{S}), one can recover as special cases the network inequalities reported in Refs.~\cite{star2017,star2017map,star2021,star2022multi,starPRA,chainQIP,chain2022,treePRA,treemulti}. Further examples illustrating our construction method are presented in Appendix~\ref{exam}.

\section{quantum violations for arbitrary two-qubit mixed states} 
To investigate nonlocality in noisy networks, we model each source as emitting an arbitrary two-qubit mixed state $\rho_i=\frac{1}{4}\big(\mathbb{I}\otimes\mathbb{I}+\sum_{u=1}^{3}a_u^{(i)}\sigma_u\otimes\mathbb{I}+\sum_{v=1}^{3}b_v^{(i)}\mathbb{I}\otimes\sigma_l+\sum_{u,v=1}^{3}\mathbf{t}_{uv}^{(i)}\sigma_u\otimes\sigma_v\big)$, with $\sigma_u$($u$=1,2,3) the Pauli operators and  $\mathbf{t}_{uv}^{(i)}=\text{Tr}[\rho_{i}(\sigma_u\otimes\sigma_v)]$ the entries of the correlation matrix $T_{\rho_{i}}$ of $\rho_{i}$.  
Each party performs  dichotomic measurements that are separable across the qubit systems. 

	For any ensemble of two-qubit mixed states $\rho=\bigotimes_{i=1}^{M}\rho_{i}$, the maximal quantum value of Eq. (\ref{S}) obtained using local qubit observables satisfies
\begin{eqnarray}\label{T3}
	S^{\text{max}}(\rho)\leqslant\prod_{i\in\mathcal{I}}\Big[\mathcal{B}^{\text{max}}_i(\rho_i)\Big]^{\frac{1}{l}}\prod_{u\in[M]\backslash\mathcal{I}}t_{u,0}^{\frac{1}{l}}\label{T12},
\end{eqnarray}
where $\mathcal{B}^{\text{max}}_i(\rho_i)$ denotes the  maximal quantum value of $\mathcal{B}_i$ that can be achieved for the shared two-qubit state $\rho_i$, and $t_{i,0}$ is the largest singular value of $T_{\rho_{i}}$.
 
The upper bound in Eq.~(\ref{T3}) is achieved when the operators acting on each peripheral source $\rho_i (i\in \mathcal{I})$ both maximize $\mathcal{B}_{i}(\rho_i)$ and satisfy either $\text{Rank}(\mathbf{X})=1$ or, for some $i\in\mathcal{I}$, $x_{j,i}=0$ for all $j\in[k]$. Here $\mathbf{X}=(x_{j,i})$ with $x_{j,i}=\langle B^{(i)}_j\Delta^{(i)}_j\rangle_{\rho_{i}}\geqslant0$. Meanwhile, the operators acting on intermediate source $\rho_i (i\in[M]\backslash\mathcal{I})$ should enable $\langle A_{j}^{(i)}B_{j}^{(i)}\rangle_{\rho_{i}}= t_{i,0}$. The detail proof is provided in Appendix \ref{appe21}. An immediate consequence is that to attain the maximal violation, it is sufficient to measure the systems connected by intermediate sources in the $\sigma_3$ basis.

\section{Resistance to noise}\label{noise}
To analyze the noise robustness of the inequalities derived in this work, we assume that each source emits a Werner state, $\rho_i=v_i|\phi_i\rangle\langle\phi_i|+\frac{1-v_i}{4}\mathbb{I}$, with visibility $v_i\in[0,1]$. The pure state  $|\phi_i\rangle=a_i|00\rangle+b_i|11\rangle$ is a generalized EPR state, where $a_i,b_i$ are real Schmidt coefficients  satisfying $a_i^2+b_i^2=1$. 
In the two inputs case with $x_i,j\in\{1,2\}$, the CHSH inequality is considered for each peripheral source such that $M_{x_i,j}=\frac{1}{2}(-1)^{x_i\cdot j}$ and $(\prod_{i\in\Gamma}\beta_i)^\frac{1}{l}$=1.  
From Eq.~(\ref{T3}), the Bell inequality (\ref{S}) is violated whenever the visibilities satisfy
\begin{eqnarray}\label{N12}
	\prod_{i=1}^Mv_i&>&\frac{1}{\prod_{i\in\mathcal{I}}\sqrt{1+4a_i^2b_i^2}}.
\end{eqnarray} 
 
We further extend the analysis to the case of $k$ inputs per party with $x_i,j\in[k]$. In this setting, the chained Bell inequality is considered for each peripheral source, which leads to  $M_{x_i,j}=\frac{1}{2}[\delta_{x_i,j}+(1-2{\delta_{j,k}})\delta_{x_i,j+1(\text{mod}k)}]$ and $(\prod_{i\in\Gamma}\beta_i)^\frac{1}{l}=k-1$.  
Using Eq.~(\ref{T3}), we obtain $S^{\text{max}}(\rho)\leqslant k\cos\frac{\pi}{2k}\prod_{i\in[M]}v_i^{\frac{1}{l}}$, where the bound is tight whenever the largest singular value of $T_{\rho_{i}}(i\in\mathcal{I})$ is degenerate. For instance, equality holds when all sources emit the Werner states that mixed the maximally entangled state $|\phi_i\rangle=\frac{1}{\sqrt{2}}(|00\rangle+|11\rangle)$ with white noise $\mathbb{I}$. In this case, the nonlocality in quantum network can be revealed when the visibilities satisfy
\begin{eqnarray}\label{N22}
	\prod_{i\in[M]}v_i&>&\Big(\frac{k-1}{k\cos\frac{\pi}{2k}}\Big)^l. 
\end{eqnarray}
When all sources produce the same Werner state  with visibility $v$ and $|\phi_i\rangle=\frac{1}{\sqrt{2}}(|00\rangle+|11\rangle)$, the critical visibility per source simplifies to $\Big(\frac{k-1}{k\cos\frac{\pi}{2k}}\Big)^{\frac{l}{M}}$.

\section{Distinguishing general network topologies}\label{DN}
One can also exploit our inequalities as a tool to distinguish the network topologies. Consider two networks $\mathcal{N}_1$ and $\mathcal{N}_2$ with the same size and different maximal number of leaf nodes (MLNs) $l_{\mathcal{N}_1}$ and $l_{\mathcal{N}_2}$. Without loss of generality, we apply inequalities (\ref{S}) and (\ref{SQ}) under the assumption of two inputs per party. In this case, the left-hand side of both inequalities reduces to the form
\begin{eqnarray}\label{S2}
	\mathcal{S}=|I_1(N,l)|^{\frac{1}{l}}+|I_2(N,l)|^{\frac{1}{l}},
\end{eqnarray} 
where $I_j(N,l)=\Big\langle\prod_{u\in\overline{\Gamma}}A_j^{u}\prod_{i\in\Gamma}\Delta_j^{i}\Big\rangle$, 
with $\Delta_j^{i}=\frac{1}{2}(A_2^{i}+(-1)^jA_1^{i})$. The classical bound of $\mathcal{S}$ is 1, while the optimal quantum value $(\mathcal{S})_Q^{\text{opt}}$ is $\sqrt{2}$. Quantum correlations in network $\mathcal{N}_1$ can be revealed by violating $\mathcal{S}_{\mathcal{N}_1}\leqslant1$, and bounded by the inequality $\mathcal{S}_{\mathcal{N}_1}\leqslant\sqrt{2}$. The same holds for network $\mathcal{N}_2$.  
 
For two same-size networks with different MLNs $l_{\mathcal{N}_1}> l_{\mathcal{N}_2}>1$, the network $\mathcal{N}_2$ can be distinguished from $\mathcal{N}_1$ whenever quantum correlations achievable in network $\mathcal{N}_2$ violate the inequality $\mathcal{S}_{\mathcal{N}_1} \leqslant \sqrt{2}$.
See Appendix \ref{appe31} for the proof. 

These networks can generate distinct sets of nonlocal correlations through local measurements, implying that network structures can be distinguished in a device-independent way. Next, we consider the noisy networks where each source emits Werner state, $\rho=v|\phi\rangle\langle\phi|+\frac{1-v}{4}\mathbb{I}$, with $|\phi\rangle=\frac{1}{\sqrt{2}}(|00\rangle+|11\rangle)$. 
From Sec.~\ref{noise}, the nonlocality can be revealed if  
\begin{eqnarray}\label{D2}
	v>\Big(\frac{1}{\sqrt{2}}\Big)^{\frac{l}{M}}.
\end{eqnarray} 
Eq.~(\ref{D2}) shows that the number of sources and the MLN in different network topologies influence the network's ability to generate quantum correlations. Specifically, the larger the ratio $\frac{l}{M}$, the more readily the network can produce quantum nonlocal correlations. 
This suggests that, among networks of the same size, those with larger leaf nodes exhibit a greater capacity to generate quantum nonlocal correlations.
 \begin{figure}[t!]
 	\resizebox{4.2cm}{4.7cm}{\includegraphics{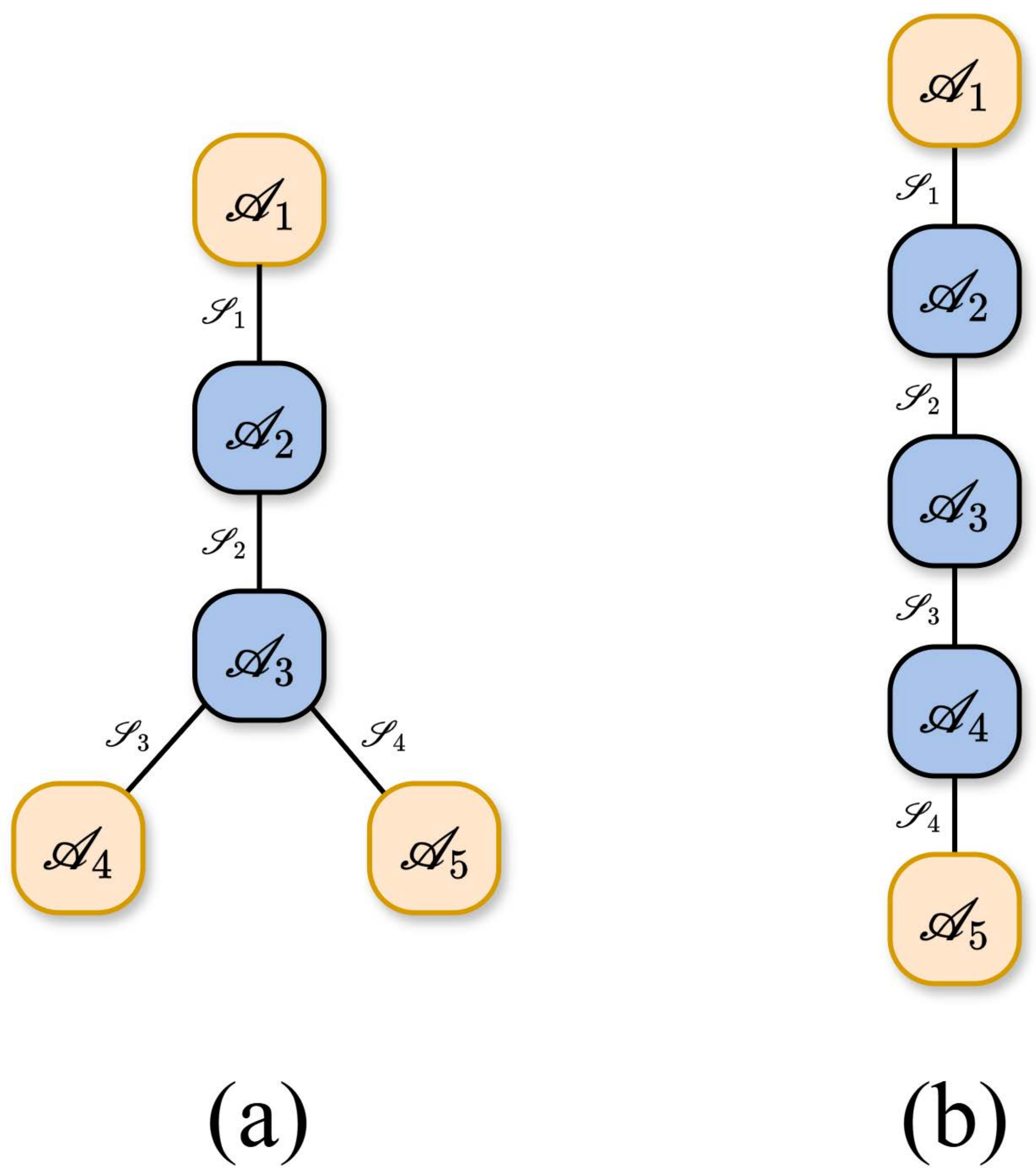}}
 	\caption{\footnotesize\justifying Two five-party, four-source networks ($N=5, M=4$). (a). Tree network with 3 leaf nodes ($l$=3). (b). Chain network with 2 leaf nodes ($l$=2).}
 	\label{D1}
 \end{figure}

To illustrate our results, we consider the two small-scale networks shown in Fig.~\ref{D1}. These networks can be distinguished by violating $\mathcal{S}_{\mathsf{(a)}}\leqslant\sqrt{2}$ using  quantum correlations generated in the network of Fig.~\ref{D1}(b) (i.e., $\mathcal{S}_{\mathsf{(b)}}\leqslant\sqrt{2}$). 
When all sources produce Werner states, Eq.~(\ref{D2}) implies that for $v\in[0.8123,0.8706]$, quantum nonlocal correlations can be detected in the network of Fig. \ref{D1}(a), but not in that of Fig. \ref{D1}(b). Therefore, the two networks can be distinguished.

Note that the two networks in Fig. \ref{D1} cannot be distinguished by the inequality proposed in \cite{Luo2018}, as they have the same number of mutually unconnected parties \cite{witness}. 
Similarly, our inequalities are not applicable for distinguishing networks with the same MLN. This observation suggests that effectively distinguishing network structures may require the combination of different types of inequalities.
  
\section{Conclusions} 
In summary, we provided a family of explicit nonlinear network  Bell inequalities with multiple inputs, exploiting their structural connection to bipartite FCBIs. 
The construction relies on the MLN, a graph-theoretic quantity computable in linear time. Since the MLN never exceeds the maximum number of mutually unconnected parties, our inequalities possess a simpler structural form than those in Ref.~\cite{Luo2018}, substantially reducing the experimental effort required to observe their violation.  
Using the elegant SOS approach, we analytically determined the optimal quantum violation of these network inequalities, explicitly expressed as a function of the quantum bounds of the FCBIs associated with peripheral sources. Our results showed that measurement strategies achieving optimal FCBI violations can be extended to general networks. The optimal quantum value remains achievable even when intermediate sources emit classical states, provided that the largest singular value of their correlation matrix equals one. Moreover, we derived the upper bound on maximal quantum violation for arbitrary ensembles of two-qubit mixed states under separable measurements, and identified the conditions required to achieve it.

The main strengths of our inequalities lie in their scalability to networks with increasing numbers of nodes and measurement inputs per party, as well as in the possibility of achieving maximal quantum violation using separable measurements, which offers a significant experimental advantage.  
Moreover, our inequalities complement existing topological witness schemes by distinguishing same-sized networks with equal numbers of mutually unconnected parties, which cannot be distinguished by the method in Ref.~\cite{witness}.
 
{\it Acknowledgements:} This work is supported by the National Natural Science Foundation of China (Grants No. 62171056, No. 62571060, and No. 62220106012).

\appendix
\begin{widetext}
\section{Proof of the inequality~(\ref{S}) and (\ref{SQ})}

In this section, we present detailed proofs of the classical bounds and optimal quantum violations for the network Bell inequalities presented in the main text. We start by introducing bipartite full-correlation Bell inequalities and their quantum optimal values, which serve as the foundation for analyzing the corresponding quantities in the proposed nonlinear network Bell inequalities.
  
\subsection{Bipartite full-correlation Bell inequalities} \label{appe11}
In the standard bipartite Bell scenario, a single source distributes systems to two spatially separated parties, $\mathscr{A}$ and $\mathscr{B}$. Each party performs a dichotomic measurement, denoted by $A_x$ for $\mathscr{A}$ and $B_y$ for $\mathscr{B}$, corresponding to inputs $x\in\mathcal{X}$ and $y\in\mathcal{Y}$, respectively. 
For convenience, the measurement outcomes are denoted as $A_x, B_y=\pm1$. Any bipartite full-correlation Bell inequality (FCBI) takes the form \cite{star2017map}
\begin{eqnarray}\label{FB}
	\mathcal{B}_{\mathbf{M}}=\sum_{x\in\mathcal{X}}\sum_{y\in\mathcal{Y}}M_{x,y}\langle A_{x}B_{y}\rangle\leqslant \beta,
\end{eqnarray}
where $\beta$ is the classical bound, and $M_{x,y}$ are real numbers. The term  $\langle A_{x}B_{y}\rangle$ represents	 the expectation value of the measurements outcomes of parties $\mathscr{A}$ and $\mathscr{B}$. The Bell inequality $\mathcal{B}_{\mathbf{M}}$ is fully characterized by a coefficient matrix $\mathbf{M} = (M_{x,y}) \in \mathbb{R}^{|\mathcal{X}| \times |\mathcal{Y}|}$. As a specific example, the CHSH inequality $\mathcal{B}_\text{CHSH}$ is obtained by setting $M_{x,y} = \frac{1}{2}(-1)^{xy}$, where $x, y \in \{1,2\}$. To compute the classical bound $\beta$, we rewrite the Eq. (\ref{FB}) to
\begin{eqnarray}\label{FBII}
	\mathcal{B}_{\mathbf{M}}=\sum_{y\in\mathcal{Y}}\langle \Delta_yB_{y} \rangle,
\end{eqnarray}
where $\Delta_y=\sum_{x\in\mathcal{X}}M_{x,y} A_{x}$. Setting $B_y=\text{sign}(\Delta_y)$, we have  
\begin{eqnarray}\label{full}
	\beta=\max_{\{A_x=\pm1\}_{x\in\mathcal{X}}}\sum_{y\in\mathcal{Y}}|\Delta_y|.
\end{eqnarray}

To derive the optimal quantum value of the Bell inequality $\mathcal{B}_{\mathbf{M}}$, we use the elegant sum-of-squares (SOS) approach introduced in \cite{star2021}. Consider that $(\mathcal{B}_{\mathbf{M}})_Q\leqslant \delta_{\mathbf{M}}$, where $\delta_{\mathbf{M}}$ is the upper bound of $(\mathcal{B}_{\mathbf{M}})_Q$. Equivalently, there exists a positive semi-definite operator $\gamma_{\mathcal{B}_{\mathbf{M}}}\geqslant0$ such that $\langle\gamma_{\mathcal{B}_{\mathbf{M}}}\rangle_Q=-(\mathcal{B}_{\mathbf{M}})_Q+\delta_{\mathbf{M}}$. The optimal quantum value of $(\mathcal{B}_{\mathbf{M}})_Q$ is attained when $\langle\gamma_{\mathbf{B}_{\mathbf{M}}}\rangle_Q=0$. The existence of such operator $\gamma_{\mathcal{B}_{\mathbf{M}}}$ can be proved by considering a set of positive operators $\{L_y, y\in\mathcal{Y}\}$ which are linear functions of the observables $\{A_x\}$, $\{B_y\}$, such that
\begin{eqnarray}\label{Q1}
	\gamma_{\mathcal{B}_{\mathbf{M}}}=\sum_{y\in\mathcal{Y}}\frac{\omega_y}{2}L_y^{\dagger}L_y,
\end{eqnarray} 
where the operators $L_y$ $(\forall y)$ and the quantities $\omega_y$ are given as follows
\begin{eqnarray}\label{Q2}
	L_y\vert\psi\rangle&=&\frac{\Delta_y}{\omega_y}-B_y,\label{Q21}\\ \omega_y&=&\Vert\Delta_{y}\vert\psi\rangle\Vert_2,\label{Q22}
\end{eqnarray} 
where $\Vert\cdot\Vert_2$ is the Frobenious norm, given by $\Vert\mathcal{O}\vert\psi\rangle\Vert_2=\sqrt{\langle\psi\vert\mathcal{O}^{\dagger}\mathcal{O}\vert\psi\rangle}$.

Substituting Eq. (\ref{Q21}) and Eq. (\ref{Q22}) into  Eq. (\ref{Q1}) and noting that $A_i^{\dagger}A_i=B_i^{\dagger}B_i=\mathbb{I}$, we get
$\langle\gamma_{\mathcal{B}_{\mathbf{M}}}\rangle_Q=\sum_y\omega_y-(\mathcal{B}_{\mathbf{M}})_Q$. Since $\langle\gamma_{\mathcal{B}_{\mathbf{M}}}\rangle_Q\geqslant0$, we have the optimal value of $(\mathcal{B}_{\mathbf{M}})_Q$ when $\langle\gamma_{\mathcal{B}_{\mathbf{M}}}\rangle_Q=0$. Then we obtain that
\begin{eqnarray}\label{QF}
	(\mathcal{B}_{\mathbf{M}})_Q^{\text{opt}}=\text{max}\Big(\sum_{y\in\mathcal{Y}}\omega_y\Big).
\end{eqnarray} 
 
\subsection{Proof of the inequality~(\ref{S})} \label{appe12}

In this subsection, we provide the full proof of the classical bound presented in Eq.~(\ref{S}). As a preliminary step, we recall an inequality established in \cite{maximal2023}:

Let $\mathbf{X} = (x_{j,i}) \in \mathbb{R}^{p \times q}$ be a non-negative matrix where $x_{j,i}\geqslant 0$; then
\begin{eqnarray}\label{C-S}
	\sum_{j=1}^{p}\left(\prod_{i=1}^{q}x_{j,i}\right)^{\frac{1}{q}}\leqslant \prod_{i=1}^{q}\left(\sum_{j=1}^{p}x_{j,i}\right)^{\frac{1}{q}},
\end{eqnarray}
where the equality holds if and only if either $\text{Rank}(\mathbf{X})=1$, or for some $i\in[q]$, $x_{j,i}=0$ for all $j\in[p]$.

Assume that Eq. (\ref{LHV}) holds.    
We use $A_j^{u}$ to represent $A_{x_u=j}$ and denote $\Delta_j^{i}=\sum_{x_i}M_{x_i,j}A_{x_i}$ for all $i\in\Gamma$. The term in Eq. (\ref{I}) is rewritten as  
\begin{eqnarray}\label{I1}
	I_j(N,l)=\Big\langle\prod_{u\in\overline{\Gamma}}A_j^{u}\prod_{i\in\Gamma}\Delta_j^{i}\Big\rangle.
\end{eqnarray} 
Moreover, we write $\big\langle A_j^{u}\big\rangle_{\Lambda_u}=\sum_{a_u=0}^1(-1)^{a_u}p(a_u|x_u,\Lambda_u)$.
Since $|\big\langle A_j^{u}\big\rangle_{\Lambda_u}|\leqslant 1$, using Eqs. (\ref{LHV}) and (\ref{I1}), we have 
\begin{eqnarray}\label{I2}
	|I_j(N,l)|&=&\Big|\int \prod_{v=1}^{M}d\lambda_v q(\lambda_{v})\prod_{u\in\overline{\Gamma}}\big\langle A_j^{u}\big\rangle_{\Lambda_u}\prod_{i\in\Gamma}\big\langle \Delta_j^{i}\big\rangle_{\Lambda_i}\Big|\\&\leqslant&\int \prod_{v=1}^{M}d\lambda_v q(\lambda_{v})\prod_{u\in\overline{\Gamma}}\Big|\big\langle A_j^{u}\big\rangle_{\Lambda_u}\Big|\cdot\prod_{i\in\Gamma}\Big|\big\langle \Delta_j^{i}\big\rangle_{\Lambda_i}\Big|\\&\leqslant&\int \prod_{v=1}^{M}d\lambda_v q(\lambda_{v})\prod_{i\in\Gamma}\Big|\big\langle \Delta_j^{i}\big\rangle_{\Lambda_i}\Big|\\&\leqslant& \prod_{i\in\Gamma}\int  d\Lambda_i q(\Lambda_{i})\Big|\big\langle \Delta_j^{i}\big\rangle_{\Lambda_i}\Big|.
\end{eqnarray}  
Following the notation in the main text, we denote by $\mathscr{A}_i$ the leaf node connected to peripheral source $\mathscr{S}_i$, with its associated hidden variable set defined as $\Lambda_i=\{\lambda_i\}$. Thus, we have
\begin{eqnarray} \label{SL1}
	(\mathcal{S})_L&=&\sum_{j=1}^k|I_j(N,l)|^\frac{1}{l}\\&\leqslant&\sum_{j=1}^k\Big|\prod_{i\in\Gamma}\int  d\lambda_i q(\lambda_i)\big|\big\langle \Delta_j^{i}\big\rangle_{\lambda_i}\big|\Big|^\frac{1}{l}\\&\leqslant&\Big[\prod_{i\in\Gamma}\int  d\lambda_i q(\lambda_i)\sum_{j=1}^k\big|\big\langle \Delta_j^{i}\big\rangle_{\lambda_i}\big|\Big]^\frac{1}{l},
\end{eqnarray} 
where the last inequality follows from Eq.~(\ref{C-S}). Substituting the classical bound (\ref{full}) of bipartite FCBI into the integrand, and noting that $\forall i: \int d\lambda_{i}q(\lambda_{i})=1$, we obtain 
\begin{eqnarray} \label{SL2}
	(\mathcal{S})_L&\leqslant&\Big[\prod_{i\in\Gamma}\big(\mathcal{B}_i\big)_L\Big]^\frac{1}{l},  
\end{eqnarray} 
where $\mathcal{B}_i$ is the bipartite FCBI associated with  peripheral source $\mathscr{S}_i$, which is fully characterized by a real coefficient matrix $\mathbf{M}^{(i)}=(M_{x_i,j})$.

\subsection{Proof of the inequality (\ref{SQ})}\label{appe13}
In this subsection, we prove the optimal quantum value presented in Eq.~(\ref{SQ}). 

In quantum theory, each independent source $\mathscr{S}_i$ emits a bipartite quantum state  $\rho_i=|\psi\rangle_i\langle\psi|_i$. The overall quantum state of network is thus given by $\rho=|\psi\rangle\langle\psi|=\bigotimes_{i=1}^{M}\rho_{i}$, where $|\psi\rangle=\bigotimes_{i=1}^{M}|\psi\rangle_i$.  To derive the optimal quantum value of $(\mathcal{S})_Q$, we use the similar elegant SOS approach as stated in Sec. \ref{appe11}. 

Let us consider that $(\mathcal{S})_Q\leqslant \delta$, where $\delta$ is clearly
the upper bound of $(\mathcal{S})_Q$. We show that there exists a positive semi-definite  operator $\gamma_{\mathcal{S}}$ satisfying $\langle\gamma_{\mathcal{S}}\rangle_Q=-(\mathcal{S})_Q+\delta$. The operator is
constructed from a set of positive operators $\{L_j, j=1,2,\ldots, k\}$, each being a polynomial function of the observables $A_{x_1},A_{x_2},\ldots,A_{x_N}$, such that 
\begin{eqnarray} \label{R}
	\langle\gamma_{\mathcal{S}}\rangle_Q=\sum_{j=1}^k\frac{\omega_{j}^{\frac{1}{l}}}{2}\langle \psi|L_j^{\dagger}L_j|\psi\rangle,
\end{eqnarray} 
where is $\omega_{j}$ a positive number with $\omega_{j}=\prod_{i\in\Gamma}\omega_{j}^{A_{i}}$.   The operators $L_j$ are defined as
\begin{eqnarray} \label{L}
	\big\vert L_j|\psi\rangle\big\vert&=&\Big\vert\prod_{i\in\Gamma}\frac{\Delta_{j}^{i}}{\omega_{j}^{A_{i}}}|\psi\rangle\Big\vert^\frac{1}{l}-\Big\vert \prod_{u\in\overline{\Gamma}}A_j^{u}\vert\psi\rangle\Big\vert^\frac{1}{l},\label{L1}\\ \omega_{j}^{A_{i}}&=&\Vert\Delta^{i}_{j}\vert\psi\rangle\Vert_2, \label{L2}
\end{eqnarray}
where $\Delta_j^{i}=\sum_{x_i}M_{x_i,j}A_{x_i}$. From this construction, it follows that
\begin{eqnarray} \label{EL}
	\langle\psi|L_j^{\dagger}L_j|\psi\rangle=2-\frac{2}{(\omega_j)^{\frac{1}{l}}}\Big\vert\langle\psi|\prod_{i\in\Gamma}\Delta_{j}^{i}\prod_{u\in\overline{\Gamma}}A_j^{u}|\psi\rangle\Big\vert^\frac{1}{l}, 
\end{eqnarray}
where we have used Eqs. (\ref{L1}) and (\ref{L2}), together with  $(A_j)^2=\mathbb{I}$. Substituting Eq. (\ref{EL}) into Eq. (\ref{R}), we have
\begin{eqnarray} \label{gamma}
	\langle\gamma_{\mathcal{S}}\rangle_Q=\sum_j\omega_j^\frac{1}{l}-(\mathcal{S})_Q. 
\end{eqnarray}
Since $\langle\gamma_{\mathcal{S}}\rangle_Q\geqslant0$, the optimal quantum value of $(\mathcal{S})_Q$ is achieved when $\langle\gamma_{\mathcal{S}}\rangle_Q=0$, which yields
\begin{eqnarray}\label{SOQ}
	(\mathcal{S})_Q^{\text{opt}}=\text{max}\Big(\sum_{j=1}^k\omega_j^\frac{1}{l}\Big).
\end{eqnarray} 
Noting that $\omega_{j}=\prod_{i\in\Gamma}\omega_{j}^{A_{i}}$ and combining Eq.~(\ref{C-S}), we obtain
\begin{eqnarray}\label{QS}
	(\mathcal{S})_Q^{\text{opt}}&\leqslant&\prod_{i\in\Gamma}\Big[\sum_{j=1}^k\omega_{j}^{A_{i}}\Big]^\frac{1}{l}\label{QS1}\\&\leqslant&\Big[
	\prod_{i\in\Gamma}\big(\mathcal{B}_i\big)_Q^{\text{opt}}\Big]^{\frac{1}{l}},\label{QS2}
\end{eqnarray}  
where the second inequality follows from Eq.~(\ref{QF}), which gives the optimal quantum value of the bipartite FCBI $\mathcal{B}$ characterized by  coefficient matrix $\mathbf{M}=(M_{x,y})$. Quantity $(\mathcal{B}_i)^{\text{opt}}_Q$ denotes the optimal quantum violation of the inequality $\mathcal{B}_i$, fully characterized by $\mathbf{M}^{(i)}=(M_{x_i,j})$.
The upper bound in Eq.~(\ref{QS2}) can be achieved if and only if the operators realizing $\big(\mathcal{B}_{i}\big)_Q^{\text{opt}}$ satisfy the equality condition of Eq.~(\ref{C-S}) with $x_{j,i}=\omega_{j}^{A_{i}}$. In addition, for every $j$, the condition $\langle \psi|L_j^{\dagger}L_j|\psi\rangle=0$ must hold, which is equivalent to $\langle\prod_{i\in\Gamma}\frac{\Delta_{j}^{i}}{\omega_{j}^{A_{i}}}\prod_{u\in\overline{\Gamma}}A_j^{u}\rangle_{\rho}=1$, as follows from Eq.~(\ref{EL}).  
\subsection{Conditions for achieving upper bound in inequality (\ref{SQ}) with separable measurements}\label{appe14}
In what follows, we make a assumption that intermediate parties perform separable measurements. In our framework, each intermediate party receives multiple particles, while each leaf node receives a single particle. Intermediate parties perform separable measurements on each of their subsystems. Denote by $A_{x_\square}^{(i)}$ and $B_{x_\triangle}^{(i)}$ the observables acting on the two subsystems of state  $\rho_i$ in the network, with $\mathscr{A}_{\square}$ and $\mathscr{A}_{\triangle}$
connected via $\rho_i$. Under this assumption, the measurement of an intermediate party $\mathscr{A}_i$ $(i\in\overline{\Gamma})$ generally factorizes as
\begin{eqnarray}\label{separa}
 A_{x_i}=\prod_{u\in U\subset\mathcal{I}_i}A^{(u)}_{x_i}\prod_{v\in \mathcal{I}_i\backslash U}B^{(v)}_{x_i},
 \end{eqnarray}
where $\mathcal{I}_i$ is the set of indices of sources associated with party $\mathscr{A}_i$. We define $\mathcal{I}=\cup_{i\in{\Gamma}}\mathcal{I}_i$ as the set of indices of all peripheral sources, and $\overline{\mathcal{I}}=[M]\backslash\mathcal{I}$ as the set of indices of all intermediate sources. As specified in the main text,  $\mathcal{I}=\Gamma$. For a leaf node $\mathscr{A}_i$$(i\in\Gamma)$, the measurement reduces to that on its single subsystem, namely  $A_{x_i}=A^{(i)}_{x_i}$.
 
Under this assumption, we now examine the conditions required to achieve the upper bound (\ref{QS2}).

First, to make the equality in (\ref{QS2}) hold, the local observables $\{A_{x_i}^{(i)}, x_i\in|\mathcal{X}_i|\}$ and $\{B_{j}^{(i)}, j\in[k]\}$ acting on the peripheral source $\mathscr{S}_i$ $(i\in\mathcal{I})$ should optimize the value of 
$\big(\mathcal{B}_{i}\big)_Q^{\text{opt}}$. 

Then, to saturate inequality (\ref{QS1}), the operators $\{A_{x_i}^{(i)}, x_i\in|\mathcal{X}_i|\}$ and $\{B_{j}^{(i)}, j\in[k]\}$ also should satisfy the equality condition of Eq.~(\ref{C-S}), that is either $\text{Rank}(\mathbf{X})=1$ or, for some $i\in\mathcal{I}$, $x_{j,i}=0$ for all $j\in[k]$, where $\mathbf{X}=(x_{j,i})$ with $x_{j,i}=\omega_{j}^{A_{i}}=\Vert\Delta^{i}_{j}\vert\psi\rangle_i\Vert_2,\Delta_j^{i}=\sum_{x_i}M_{x_i,j}A_{x_i}^{(i)}$.

Finally, the optimal quantum value in Eq.~(\ref{SOQ}) will occur under the optimization condition  $\langle\gamma_{\mathcal{S}}\rangle_Q=0$. From Eqs.~(\ref{R}) and (\ref{EL}), this is equivalent to $\langle\prod_{i\in\Gamma}\frac{\Delta_{j}^{i}}{\omega_{j}^{A_{i}}}\prod_{u\in\overline{\Gamma}}A_j^{u}\rangle_{\rho}=1$, for all $j\in[k]$. According to Eq.~(\ref{separa}), this condition becomes
\begin{eqnarray}\label{condi}
	\langle\prod_{i\in\Gamma}\frac{\Delta_{j}^{i}}{\omega_{j}^{A_{i}}}\prod_{u\in\overline{\Gamma}}A_j^{u}\rangle_{\rho}=\prod_{i\in\mathcal{I}}\langle\frac{\Delta_{j}^{(i)}}{\omega_{j}^{A_{i}}}B_{j}^{(i)}\rangle_{\rho_i}\prod_{u\in\overline{\mathcal{I}}}\langle A_{j}^{(u)}B_{j}^{(u)}\rangle_{\rho_u}=1.
\end{eqnarray}

From the derivation of optimal quantum violations for the bipartite FCBIs (see Sec.~\ref{appe11}), it follows that $(\mathcal{B}_i)^{\text{opt}}_Q$ will occur under the condition $\langle\gamma_{\mathcal{B}_i}\rangle_Q=0$, which is equivalent to $\langle\frac{\Delta_{j}^{(i)}}{\omega_{j}^{A_{i}}}B_{j}^{(i)}\rangle_{\rho_i}=1$ for all $i\in\mathcal{I}$. Therefore, to satisfy (\ref{condi}), it suffices to ensure that for every intermediate source $\mathscr{S}_i$ $(i\in\overline{\mathcal{I}})$, the condition $\langle A_{j}^{(i)}B_{j}^{(i)}\rangle_{\rho_{i}}=1$ holds for all inputs.

\section{Proof of the Eq.~(\ref{T3})}\label{appe21}
In this section, we derive the maximal quantum violations of Eq. (\ref{S}) obtained by arbitrary ensembles of two-qubit states.
Let all sources in Fig. \ref{fig1} emit any two-qubit mixed states $\rho_1,\rho_2,\ldots,\rho_M$, which can be expressed in terms of Pauli matrices as 
\begin{eqnarray}\label{rho}
	\rho_i=\frac{1}{4}\big(\mathbb{I}\otimes\mathbb{I}+\sum_{u=1}^{3}a_u^{(i)}\sigma_u\otimes\mathbb{I}+\sum_{v=1}^{3}b_v^{(i)}\mathbb{I}\otimes\sigma_l+\sum_{u,v=1}^{3}\mathbf{t}_{uv}^{(i)}\sigma_u\otimes\sigma_v\big),
\end{eqnarray}
where $T_{\rho_{i}}=(\mathbf{t}_{uv}^{(i)})$ with $\mathbf{t}_{uv}^{(i)}=\text{Tr}[\rho_{i}(\sigma_u\otimes\sigma_v)]$ is the correlation matrix of the state $\rho_{i}$, and $\sigma_u (u=1,2,3)$ are the Pauli matrices $\sigma_x$,  $\sigma_y$, $\sigma_z$, respectively. Denote $1\geqslant t_{i,0}\geqslant t_{i,1}\geqslant t_{i,2}\geqslant 0$ the three largest singular values of $T_{\rho_{i}}$.
In our framework, intermediate nodes receive multiple qubits, while the leaf nodes receive only one qubit. Denote by $\mathcal{I}_i$ the set of indices of all sources associated with party $\mathscr{A}_i$, and define $\mathcal{I}=\cup_{i\in\Gamma}\mathcal{I}_i$ as the set of indices of all peripheral sources.

We consider the  measurements that are separable across the qubit systems. According to Eq.~(\ref{separa}), the Eq.~(\ref{I1}) can be rewritten as
\begin{eqnarray}\label{T}
	I_j(N,l)=\prod_{i\in\mathcal{I}}\Big\langle B^{(i)}_j\Delta^{(i)}_j\Big\rangle_{\rho_i}\cdot\prod_{u\in[M]\backslash\mathcal{I}}\Big\langle A^{(u)}_j B^{(u)}_j\Big\rangle_{\rho_u},
\end{eqnarray}
where $\langle B^{(i)}_j\Delta^{(i)}_j\rangle$ are the full-correlation observables defined in Eq. (\ref{FBII}). Substituting Eq.~(\ref{T}) into Eq.~(\ref{S}), and noting that $\langle A_j^{(u)}B_j^{(u)}\rangle\leqslant t_{u,0}$ for all $j\in[k]$ (with saturation when $A_j^{(u)}=B_j^{(u)}=\sigma_3$), we obtain
\begin{eqnarray}\label{T1}
	S^{\text{max}}(\rho)&\leqslant&\sum_{j=1}^{k}\prod_{i\in\mathcal{I}}\Big|\Big\langle B^{(i)}_j\Delta^{(i)}_j\Big\rangle_{\rho_i}\Big|^{\frac{1}{l}}\prod_{u\in[M]\backslash\mathcal{I}}t_{u,0}^{\frac{1}{l}}\label{T11}\\&\leqslant&\prod_{i\in\mathcal{I}}\Big[\mathcal{B}^{\text{max}}_{\mathcal{M}^{(i)}}(\rho_i)\Big]^{\frac{1}{l}}\prod_{u\in[M]\backslash\mathcal{I}}t_{u,0}^{\frac{1}{l}}\label{T12},
\end{eqnarray}
where $\mathcal{B}_{\mathcal{M}^{(i)}}(\rho_i)=\sum_{j=1}^k\Big\langle B^{(i)}_j\Delta^{(i)}_j\Big\rangle_{\rho_i}$, and we assume that $\langle B^{(i)}_j\Delta^{(i)}_j\rangle\geqslant0$ because the non-negativity condition can be ensured through appropriate orientation adjustments of operator $B_{j}^{(i)}$. The set of observables is not guaranteed to maximize $\mathcal{B}_{\mathcal{M}^{(i)}}(\rho_i)$, which explains the inequality in (\ref{T12}). The equalities in Eqs.~(\ref{T11}) and (\ref{T12}) hold when $\langle A_{j}^{(u)}B_{j}^{(u)}\rangle_{\rho_{u}}= t_{u,0}$ for all $j\in[k]$ and 
the operators that maximize $\mathcal{B}_{\mathcal{M}^{(i)}}(\rho_i)$ also satisfy the equality condition of Eq.~(\ref{C-S}), with $x_j^i=\langle B^{(i)}_j\Delta^{(i)}_j\rangle_{\rho_{i}}\geqslant0$.

\section{Examples}\label{exam}
In what follows, we randomly select a six-party, six-source network ($N$=6, $M$=6) with three leaf nodes (see Fig.~\ref{exp1}) to illustrate our construction method. The leaf nodes are  $\mathscr{A}_1,\mathscr{A}_3,\mathscr{A}_5$, while the peripheral sources are $\mathscr{S}_1,\mathscr{S}_3,\mathscr{S}_5$, implying  $\Gamma=\mathcal{I}=\{1,3,5\}$. For different measurement configurations, we provide the corresponding network Bell inequalities, their maximal quantum violations, and the measurement schemes required to achieve these violations.

Example 1. (Two-input case) Each party has two inputs and outputs. In this setting, the CHSH inequality is considered for each peripheral source such that $M_{x_i,j}=\frac{1}{2}(-1)^{x_i\cdot j}$ and $(\prod_{i\in\Gamma}\beta_i)^\frac{1}{l}=1$.
Thus, after concretizing the inequalities (\ref{S}) and (\ref{I}), we obtained that the quantum nonlocal correlations from the network are demonstrated by violating the following inequality
\begin{eqnarray}\label{E1}
	\mathcal{S}_{\text{exp1}}=|I_1(6,3)|^{\frac{1}{3}}+|I_2(6,3)|^{\frac{1}{3}}&\leqslant&1,
\end{eqnarray}
where $I_1(6,3)=\frac{1}{2^3}\langle (A^{1}_{2}-A^{1}_{1})A^{2}_{1}(A^{3}_{2}-A^{3}_{1})A^{4}_{1}(A^{5}_{2}-A^{5}_{1})A^{6}_{1}\rangle$ and $I_2(6,3)=\frac{1}{2^3}\langle (A^{1}_{1}+A^{1}_{2})A^{2}_{2}(A^{3}_{1}+A^{3}_{2})A^{4}_{2}(A^{5}_{1}+A^{5}_{2})A^{6}_{2}\rangle$. For the network, sharing six mixed states $\rho_1,\rho_2,\ldots,\rho_6$ provided by six independent sources, by applying Eq.~(\ref{T3}), 
the maximal quantum value of $\mathcal{S}_{\text{exp1}}$ is
\begin{eqnarray}\label{E1T}
	S_{\text{exp1}}^{\text{max}}(\rho)=\prod_{i\in\mathcal{I}}\big[\sqrt{t_{i,0}^2+t_{i,1}^2}\big]^{\frac{1}{3}}\prod_{u\in[M]\backslash\mathcal{I}}t_{u,0}^{\frac{1}{3}},
\end{eqnarray}
with $\mathcal{I}=\{1,3,5\}$ and $[M]\backslash\mathcal{I}=\{2,4,6\}$,
which is achievable by letting
\begin{eqnarray}\label{E1M}
	&&A_{x_j,j\in\{1,3,5\}}=x_j\sigma_1+(1-x_j)\sigma_3,\nonumber\\ &&A_{x_2}=\frac{t_{1,0}\sigma_3+t_{1,1}(-1)^{x_2}\sigma_1}{\sqrt{t_{1,0}^2+t_{1,1}^2}}\otimes\sigma_3\otimes\sigma_3,\nonumber\\
	&&A_{x_4}=\sigma_3\otimes\frac{t_{3,0}\sigma_3+t_{3,1}(-1)^{x_4}\sigma_1}{\sqrt{t_{3,0}^2+t_{3,1}^2}}\otimes\sigma_3\otimes\frac{t_{5,0}\sigma_3+t_{5,1}(-1)^{x_4}\sigma_1}{\sqrt{t_{5,0}^2+t_{5,1}^2}},\nonumber\\
	&&A_{x_6}=\sigma_3\otimes\sigma_3,\nonumber  
\end{eqnarray}
where $x_i\in\{1,2\}$.
Eq.~(\ref{SQ}) implies that the quantum value of inequality (\ref{E1}) is at most $\sqrt{2}$, attained by the measurements in Eq.(\ref{E1M}) with $t_{i,0}=t_{i,1}=1$ $(i\in\mathcal{I})$ on maximally entangled states. 
 
\begin{figure}[]
	\resizebox{7.4cm}{4.5cm}{\includegraphics{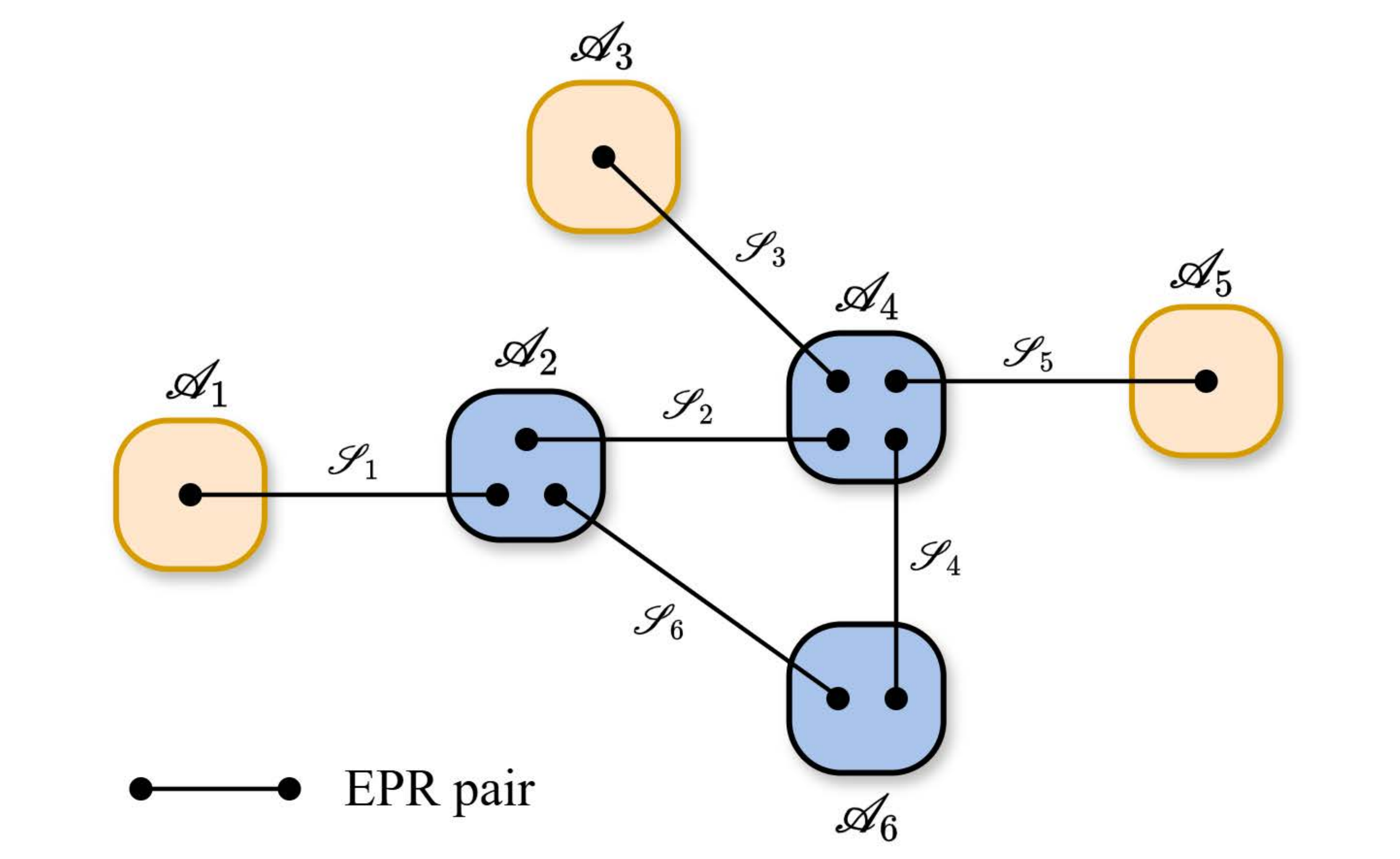}}
	\caption{\justifying\footnotesize The network consisting of six sources $\mathscr{S}_1$, $\mathscr{S}_2$, \ldots, $\mathscr{S}_6$ and six parties $\mathscr{A}_1$, $\mathscr{A}_2$, \ldots, $\mathscr{A}_6$, where $\mathscr{A}_1,\mathscr{A}_3,\mathscr{A}_5$ are the leaf nodes ($M$=6, $l$=3).}
	\label{exp1}
\end{figure}
Example 2. ($k$-input case) Each party has $k>2$ inputs. For each peripheral source, we consider the chained Bell inequality, whose classical bound and optimal quantum value have been derived as $k-1$ and $k\cos\frac{\pi}{2k}$, respectively \cite{CBIself}. Then, from Eq.~ (\ref{S}), we obtain the following inequality
\begin{eqnarray}\label{E2}
	\mathcal{S}_{\text{exp2}}=\sum_{j=1}^k|I_j(6,3)|^{\frac{1}{3}}&\leqslant&k-1,
\end{eqnarray}
where $I_j(6,3)=\frac{1}{2^3}\langle \prod_{u\in\{2,4,6\}}A_j^{u}\prod_{i\in\{1,3,5\}}(A_j^{i}+A_{j+1}^{i})\rangle$, 
with $A_{k+1}^{i}=-A_{1}^{i}$. For any ensemble of two-qubit mixed states $\rho=\otimes_{i=1}^{6}\rho_{i}$, by applying Eq.~(\ref{T3}), we have 
\begin{eqnarray}\label{E2T}
	S_{\text{exp2}}^{\text{max}}(\rho)&\leqslant&k\cos\frac{\pi}{2k}\prod_{i\in[M]}t_{i,0}^{\frac{1}{3}},
\end{eqnarray}
with saturation achieved when the largest singular value of the correlation matrices for $\rho_{1},\rho_{3},\rho_{5}$ is degenerate, and the parties perform the following observables,
\begin{align}\label{E2M}
	&A_{x_i,i\in\{1,3,5\}}=\sin\frac{({x_i}-1)\pi}{k}\sigma_1+\cos\frac{({x_i}-1)\pi}{k}\sigma_3,\nonumber\\ &A_{x_2}=(\sin\frac{(2{x_2}-1)\pi}{2k}\sigma_1+\cos\frac{(2{x_2}-1)\pi}{2k}\sigma_3)^{\otimes3},\nonumber\\
	&A_{x_4}=(\sin\frac{(2{x_4}-1)\pi}{2k}\sigma_1+\cos\frac{(2{x_4}-1)\pi}{2k}\sigma_3)^{\otimes4},\nonumber\\
	&A_{x_6}=(\sin\frac{(2{x_6}-1)\pi}{2k}\sigma_1+\cos\frac{(2{x_6}-1)\pi}{2k}\sigma_3)^{\otimes2},\nonumber
\end{align}
with $x_i\in[k]$. Employing the above observables, we find that the optimal quantum value $(\mathcal{S}_{\text{exp2}})_Q^{\text{opt}}=k\cos\frac{\pi}{2k}$ can be achieved when all sources emit two-qubit maximal entangled states.

Example 3. (Asymmetry-input case) Consider the measurement configuration where the leaf nodes $\mathscr{A}_1$ and $\mathscr{A}_3$ each have 3 inputs, while leaf node $\mathscr{A}_5$ and the intermediate nodes each have 4 inputs. In this setting, the EBI is considered for the peripheral sources $\mathscr{S}_1$ and $\mathscr{S}_3$, with classical bound $\beta_i=6$ and optimal quantum value  $(\mathcal{B}_i)_Q^{\text{opt}}=4\sqrt{3}$ \cite{EBI}. The 4-input chained Bell inequality is considered for peripheral source $\mathscr{S}_5$, with classical bound $\beta_i=3$ and optimal quantum value  $(\mathcal{B}_i)_Q^{\text{opt}}=4\cos\frac{\pi}{8}$ \cite{EBI}. Substituting these values into Eq.~(\ref{S}), we obtain the nonlinear network inequality of the form 
\begin{eqnarray}\label{E3}
	\mathcal{S}_{\text{exp3}}&=&|I_1(6,3)|^{\frac{1}{3}}+|I_2(6,3)|^{\frac{1}{3}}+|I_3(6,3)|^{\frac{1}{3}}+|I_4(6,3)|^{\frac{1}{3}}\leqslant6(\frac{1}{2})^{\frac{1}{3}},
\end{eqnarray}
where $I_j(6,3)=\langle \prod_{u\in\{2,4,6\}}A_j^{u}\prod_{i\in\{1,3,5\}}\Delta_j^{i}\rangle$. For $i\in\{1,3\}$, $\Delta_1^{i}=A_1^{i}+A_2^{i}+A_3^{i}$, $\Delta_2^{i}=A_1^{i}+A_2^{i}-A_3^{i}$, $\Delta_3^{i}=A_1^{i}-A_2^{i}+A_3^{i}$, $\Delta_4^{i}=A_1^{i}-A_2^{i}-A_3^{i}$, and
$\Delta_j^{5}=\frac{1}{2}(A_j^{5}+A_{j+1}^{5})$ satisfying $A_{5}^{5}=-A_{1}^{5}$.
 
For any ensemble of two-qubit mixed states $\rho=\otimes_{i=1}^{6}\rho_{i}$, Eq.~(\ref{T3}) gives the upper bound on maximal quantum violation, 
\begin{eqnarray}\label{E2T}
	S_{\text{exp3}}^{\text{max}}(\rho)&\leqslant& 2\big[4\sqrt{2+\sqrt{2}}\prod_{i\in\{1,3\}}\sqrt{t_{i,0}^2+t_{i,1}^2+t_{i,2}^2}\prod_{u\in\{2,4,5,6\}}t_{u,0}\big]^{\frac{1}{3}},
\end{eqnarray}
where the bound is saturated if the degeneracy of largest singular value of correlation matrix $T_{\rho_5}$ exceeds l. The observables required to achieve this bound are given by 
\begin{eqnarray}\label{E3M}
	&&A_{x_j,j\in\{1,3\}} = \frac{1}{2}x_i(2-x_j)(3-x_j)\sigma_1+ (1-x_j)^2(3-x_j)\sigma_2+\frac{1}{2}(1-x_j)(2-x_j)\sigma_3, \nonumber \\  
	&&A_{x_2}= B_{x_2}^{(1)}\otimes\sigma_3\otimes\sigma_3,   \\
	&&A_{x_4}= B_{x_4}^{(3)}\otimes B_{x_4}^{(5)}\otimes\sigma_3\otimes\sigma_3, \nonumber \\
	&&A_{x_5}= \sin\frac{({x_5}-1)\pi}{4}\sigma_1 + \cos\frac{({x_5}-1)\pi}{4}\sigma_3, \nonumber \\
	&&A_{x_6}= \sigma_3\otimes\sigma_3, \nonumber
\end{eqnarray} 
with $x_1,x_3\in\{1,2,3\}$, $x_2,x_4,x_5,x_6\in\{1,2,3,4\}$,
\begin{eqnarray}
	&&B_1^{(i)}=\frac{t_{i,0}\sigma_1+t_{i,1}\sigma_2+t_{i,2}\sigma_3}{\sqrt{t_{i,0}^2+t_{i,1}^2+t_{i,2}^2}},\nonumber
	\\&&B_2^{(i)}=\frac{t_{i,0}\sigma_1+t_{i,1}\sigma_2-t_{i,2}\sigma_3}{\sqrt{t_{i,0}^2+t_{i,1}^2+t_{i,2}^2}},\nonumber\\
	&&B_3^{(i)}=\frac{t_{i,0}\sigma_1-t_{i,1}\sigma_2+t_{i,2}\sigma_3}{\sqrt{t_{i,0}^2+t_{i,1}^2+t_{i,2}^2}},\nonumber
	\\&&B_4^{(i)}=\frac{t_{i,0}\sigma_1-t_{i,1}\sigma_2-t_{i,2}\sigma_3}{\sqrt{t_{i,0}^2+t_{i,1}^2+t_{i,2}^2}}, i=1,3, \nonumber  \\ \text{and} &&B_{x_4}^{(5)}=\sin\frac{(2{x_4}-1)\pi}{8}\sigma_1+\cos\frac{(2{x_4}-1)\pi}{8}\sigma_3.\nonumber
\end{eqnarray}
From Eq.~(\ref{SQ}), the quantum value of $\mathcal{S}_{\text{exp3}}$ is upper bounded by $2(12\sqrt{2+\sqrt{2}})^{\frac{1}{3}}$, which is achieved by applying the operators in Eq.~(\ref{E3M}) with $t_{i,0}=t_{i,1}=1$ $(i\in\mathcal{I})$ on ensembles of maximally entangled states.

It is trivial to generalize these inequalities to general $N$-party networks with multiple leaf nodes.
 
\section{Distinguishing general network topologies}\label{appe31}

In this section, we prove that the same-size networks with different MLNs can be distinguished using our inequalities. Without loss of generality, we assume that each party has two inputs. From Eq.~(\ref{SQ}), the quantum correlations in network are bounded by $\sqrt{2}$. For clarity and completeness, we formally restate the main result below.

For two same-size networks with different MLNs, $l_{\mathcal{N}_1}> l_{\mathcal{N}_2}>1$, the network $\mathcal{N}_2$ can be distinguished from $\mathcal{N}_1$ when quantum correlations achievable in  $\mathcal{N}_2$ lead to a violation of $\mathcal{S}_{\mathcal{N}_1} \leqslant \sqrt{2}$. Specifically, we demonstrate that the inequality $\mathcal{S}_{\mathcal{N}_1}\leqslant\sqrt{2}$ can be violated when $\mathcal{S}_{\mathcal{N}_2}\leqslant\sqrt{2}$ holds true.
 
Assume both networks have $N$ nodes. Let $\Gamma_{\mathcal{N}_1}$ and $\Gamma_{\mathcal{N}_2}$ denote the sets of indices of leaf nodes in network $\mathcal{N}_1$ and $\mathcal{N}_2$, respectively, with  $\Gamma_{\mathcal{N}_1}\supset \Gamma_{\mathcal{N}_2}\neq\varnothing$. From Eq.~(\ref{SQ}), the Bell inequalities for networks $\mathcal{N}_1$ and $\mathcal{N}_2$ take the form
\begin{eqnarray} 
	\mathcal{S}_{\mathcal{N}_1}&=&\Big\vert\Big\langle\prod_{u\in\overline{\Gamma}_{\mathcal{N}_1}}A_0^{u}\prod_{i\in\Gamma_{\mathcal{N}_1}}\Delta_0^{i}\Big\rangle\Big\vert^{\frac{1}{l_{\mathcal{N}_1}}}+\Big\vert\Big\langle\prod_{u\in\overline{\Gamma}_{\mathcal{N}_1}}A_1^{u}\prod_{i\in\Gamma_{\mathcal{N}_1}}\Delta_1^{i}\Big\rangle\Big\vert^{\frac{1}{l_{\mathcal{N}_1}}}\leqslant\sqrt{2},\label{APPEN1}\\
	\mathcal{S}_{\mathcal{N}_2}&=&\Big\vert\Big\langle\prod_{u\in\overline{\Gamma}_{\mathcal{N}_2}}A_0^{u}\prod_{i\in\Gamma_{\mathcal{N}_2}}\Delta_0^{i}\Big\rangle\Big\vert^{\frac{1}{l_{\mathcal{N}_2}}}+\Big\vert\Big\langle\prod_{u\in\overline{\Gamma}_{\mathcal{N}_2}}A_1^{u}\prod_{i\in\Gamma_{\mathcal{N}_2}}\Delta_1^{i}\Big\rangle\Big\vert^{\frac{1}{l_{\mathcal{N}_2}}}\leqslant\sqrt{2}.\label{APPEN2}
\end{eqnarray}	
Equation~(\ref{APPEN2}) can be alternatively expressed as:
\begin{eqnarray}\label{APPEN3}
	\mathcal{S}_{\mathcal{N}_2}=&&\Big\vert\Big\langle\prod_{u\in\overline{\Gamma}_{\mathcal{N}_1}}A_0^{u}\prod_{v\in\overline{\Gamma}_{\mathcal{N}_2}\backslash \overline{\Gamma}_{\mathcal{N}_1}}A_1^{v}\prod_{i\in\Gamma_{\mathcal{N}_2}}\Delta_0^{i}\Big\rangle\Big\vert^{\frac{1}{l_{\mathcal{N}_2}}}\nonumber\\&&+\Big\vert\Big\langle\prod_{u\in\overline{\Gamma}_{\mathcal{N}_1}}A_1^{u}\prod_{v\in\overline{\Gamma}_{\mathcal{N}_2}\backslash \overline{\Gamma}_{\mathcal{N}_1}}A_0^{v}\prod_{i\in\Gamma_{\mathcal{N}_2}}\Delta_1^{i}\Big\rangle\Big\vert^{\frac{1}{l_{\mathcal{N}_2}}}\leqslant\sqrt{2}.
\end{eqnarray} 
 
Promise that $\mathcal{S}_{\mathcal{N}_2}\leqslant\sqrt{2}$ holds true. We should prove that $\mathcal{S}_{\mathcal{N}_1}\leqslant\sqrt{2}$ can be violated. Write 
\begin{eqnarray}\label{P}
	p_0&=&2^{l_{\mathcal{N}_2}}\langle\prod_{u\in\overline{\Gamma}_{\mathcal{N}_2}}A_0^{u}\prod_{i\in\Gamma_{\mathcal{N}_2}}\Delta_0^{i}\rangle,\\
	p_1&=&2^{l_{\mathcal{N}_2}}\langle\prod_{u\in\overline{\Gamma}_{\mathcal{N}_2}}A_1^{u}\prod_{i\in\Gamma_{\mathcal{N}_2}}\Delta_1^{i}\rangle,\\
	q_0&=&2^{l_{\mathcal{N}_2}}\langle\prod_{u\in\overline{\Gamma}_{\mathcal{N}_1}}A_0^{u}\prod_{v\in\overline{\Gamma}_{\mathcal{N}_2}\backslash \overline{\Gamma}_{\mathcal{N}_1}}A_1^{v}\prod_{i\in\Gamma_{\mathcal{N}_2}}\Delta_0^{i}\rangle,\\
	q_1&=&2^{l_{\mathcal{N}_2}}\langle\prod_{u\in\overline{\Gamma}_{\mathcal{N}_1}}A_1^{u}\prod_{v\in\overline{\Gamma}_{\mathcal{N}_2}\backslash \overline{\Gamma}_{\mathcal{N}_1}}A_0^{v}\prod_{i\in\Gamma_{\mathcal{N}_2}}\Delta_1^{i}\rangle.
\end{eqnarray} 
Eq.~(\ref{APPEN2}) and Eq.~(\ref{APPEN3}) can be  further rewritten as $\frac{1}{2}(|p_0|^{\frac{1}{l_{\mathcal{N}_2}}}+|p_1|^{\frac{1}{l_{\mathcal{N}_2}}})\leqslant\sqrt{2}$ and $\frac{1}{2}(|q_0|^{\frac{1}{l_{\mathcal{N}_2}}}+|q_1|^{\frac{1}{l_{\mathcal{N}_2}}})\leqslant\sqrt{2}$, respectively. Thus, there exist $\epsilon_1,\epsilon_2\geqslant 0$ such that $|p_0|^{\frac{1}{l_{\mathcal{N}_2}}}+|p_1|^{\frac{1}{l_{\mathcal{N}_2}}}=2\sqrt{2}-\epsilon_1$ and $|q_0|^{\frac{1}{l_{\mathcal{N}_2}}}+|q_1|^{\frac{1}{l_{\mathcal{N}_2}}}=2\sqrt{2}-\epsilon_2$. By parameterizing with angles $\theta_1,\theta_2\in[0,\pi/2]$, one obtains  $|p_0|^{\frac{1}{l_{\mathcal{N}_2}}}=(2\sqrt{2}-\epsilon_1)\cos^2\theta_1$, $|p_1|^{\frac{1}{l_{\mathcal{N}_2}}}=(2\sqrt{2}-\epsilon_1)\sin^2\theta_1$, $|q_0|^{\frac{1}{l_{\mathcal{N}_2}}}=(2\sqrt{2}-\epsilon_2)\cos^2\theta_2$, $|q_1|^{\frac{1}{l_{\mathcal{N}_2}}}=(2\sqrt{2}-\epsilon_2)\sin^2\theta_2$. Then, let $\theta_1\geqslant\theta_2$, from Eq.~(\ref{APPEN1}), we obtain
\begin{eqnarray}\label{SN1}
	\mathcal{S}_{\mathcal{N}_1}&=&\Big\vert\Big\langle\prod_{u\in\overline{\Gamma}_{\mathcal{N}_1}}A_0^{u}\prod_{i\in\Gamma_{\mathcal{N}_1}}\Delta_0^{i}\Big\rangle\Big\vert^{\frac{1}{l_{\mathcal{N}_1}}}+\Big\vert\Big\langle\prod_{u\in\overline{\Gamma}_{\mathcal{N}_1}}A_1^{u}\prod_{i\in\Gamma_{\mathcal{N}_1}}\Delta_1^{i}\Big\rangle\Big\vert^{\frac{1}{l_{\mathcal{N}_1}}}\nonumber\\ &=&\Big\vert\frac{p_0+q_0}{2^{l_{\mathcal{N}_1}}}\Big\vert^{\frac{1}{l_{\mathcal{N}_1}}}+\Big\vert\frac{p_1-q_1}{2^{l_{\mathcal{N}_1}}}\Big\vert^{\frac{1}{l_{\mathcal{N}_1}}}\nonumber\\&\geqslant&\frac{1}{2}(\big||p_0|-|q_0|\big|^{\frac{1}{l_{\mathcal{N}_1}}}+\big||p_1|-|q_1|\big|^{\frac{1}{l_{\mathcal{N}_1}}})\nonumber\\&=&\frac{1}{2}\Big(\Big\vert\big((2\sqrt{2}-\epsilon_1)\cos^2\theta_1\big)^{l_{\mathcal{N}_2}}-\big((2\sqrt{2}-\epsilon_2)\cos^2\theta_2\big)^{l_{\mathcal{N}_2}}\Big\vert^{\frac{1}{l_{\mathcal{N}_1}}}\nonumber\\&&+\Big\vert\big((2\sqrt{2}-\epsilon_1)\sin^2\theta_1\big)^{l_{\mathcal{N}_2}}-\big((2\sqrt{2}-\epsilon_2)\sin^2\theta_2\big)^{l_{\mathcal{N}_2}}\Big\vert^{\frac{1}{l_{\mathcal{N}_1}}}\Big)\nonumber\\&=&\frac{1}{2}\Big(\big[\big((2\sqrt{2}-\epsilon_2)\cos^2\theta_2\big)^{l_{\mathcal{N}_2}}-\big((2\sqrt{2}-\epsilon_1)\cos^2\theta_1\big)^{l_{\mathcal{N}_2}}\big]^{\frac{1}{l_{\mathcal{N}_1}}}\nonumber\\&&+\big[\big((2\sqrt{2}-\epsilon_1)\sin^2\theta_1\big)^{l_{\mathcal{N}_2}}-\big((2\sqrt{2}-\epsilon_2)\sin^2\theta_2\big)^{l_{\mathcal{N}_2}}\big]^{\frac{1}{l_{\mathcal{N}_1}}}\Big),
\end{eqnarray}	
where in the last line is a binary function $f(\theta_1,\theta_2)$, whose extreme value is taken at $\theta_1=\theta_2$, and  $f(\theta_1,\theta_1)=\frac{1}{2}[(2\sqrt{2}-\epsilon_2)^{l_{\mathcal{N}_2}}-(2\sqrt{2}-\epsilon_1)^{l_{\mathcal{N}_2}}]^{\frac{1}{l_{\mathcal{N}_1}}}(\cos^{\frac{2l_{\mathcal{N}_2}}{l_{\mathcal{N}_1}}}\theta_1-\sin^{\frac{2l_{\mathcal{N}_2}}{l_{\mathcal{N}_1}}}\theta_1)$.
Finally, evaluating the function at endpoints yields $f(\frac{\pi}{2},0)=\frac{1}{2}[(2\sqrt{2}-\epsilon_2)^{\frac{l_{\mathcal{N}2}}{l{\mathcal{N}1}}}+(2\sqrt{2}-\epsilon_1)^{\frac{l{\mathcal{N}2}}{l{\mathcal{N}1}}}]$, which exceeds $\sqrt{2}$ for some $\epsilon_1,\epsilon_2\in[0,1)$. Therefore, $\mathcal{S}_{\mathcal{N}_1}>\sqrt{2}$ is possible, showing that inequality~(\ref{APPEN1}) can be violated.
 
\subsection{Distinguishing four-source network topologies}
To further illustrate the presented approach used to distinguish network topologies, we consider two small-scale networks (see Fig. \ref{D1}) with different MLNs, each consisting of five parties and four independent sources.
The local correlations from the networks depicted in Fig. \ref{D1} respectively admit distributions of the following forms, 
\begin{eqnarray}\label{LHV1}
	&&P^{\mathsf{(a)}}(a_1,\ldots,a_5|x_1,\ldots,x_5)\nonumber\\ \nonumber=&&\int d\lambda_1d\lambda_2d\lambda_3d\lambda_4q(\lambda_1)q(\lambda_2)q(\lambda_3)q(\lambda_4)p(a_1|x_1,\lambda_1)\\&&\times p(a_2|x_2,\lambda_1,\lambda_2)p(a_3|x_3,\lambda_2,\lambda_3,\lambda_4)p(a_4|x_4,\lambda_3)p(a_5|x_5,\lambda_4),
\end{eqnarray}
and
\begin{eqnarray}\label{LHV2}
	&&P^{\mathsf{(b)}}(a_1,\ldots,a_5|x_1,\ldots,x_5)\nonumber\\ \nonumber=&&\int d\lambda_1d\lambda_2d\lambda_3d\lambda_4q(\lambda_1)q(\lambda_2)q(\lambda_3)q(\lambda_4)p(a_1|x_1,\lambda_1)\\&&\times p(a_2|x_2,\lambda_1,\lambda_2)p(a_3|x_3,\lambda_2,\lambda_3)p(a_4|x_4,\lambda_3,\lambda_4)p(a_5|x_5,\lambda_4),
\end{eqnarray} 
which are the concretization of Eq.~(\ref{LHV}).
It is obviously known that $N=5,l=3$ for the network in Fig. \ref{D1}(a), and $N=5,l=2$ for the network in Fig. \ref{D1}(b). 
Substituting the corresponding values of $\{N,l\}$ into Eq.~(\ref{S2}), we find that the quantum correlations in these two networks respectively satisfy the following inequalities 
\begin{eqnarray}\label{DQ53}
	\mathcal{S}_{\mathsf{(a)}}=|I^{\mathsf{(a)}}_1(5,3)|^{\frac{1}{3}}+|I^{\mathsf{(a)}}_2(5,3)|^{\frac{1}{3}}&\leqslant&\sqrt{2},
\end{eqnarray}
and
\begin{eqnarray}\label{DQ52}
	\mathcal{S}_{\mathsf{(b)}}=|I^{\mathsf{(b)}}_1(5,2)|^{\frac{1}{2}}+|I^{\mathsf{(b)}}_2(5,2)|^{\frac{1}{2}}&\leqslant&\sqrt{2}.
\end{eqnarray}
Furthermore, correlations from the two networks are certified to be nonlocal by violating inequalities
\begin{eqnarray}\label{DC53}
	\mathcal{S}_{\mathsf{(a)}}=|I^{\mathsf{(a)}}_1(5,3)|^{\frac{1}{3}}+|I^{\mathsf{(a)}}_2(5,3)|^{\frac{1}{3}}&\leqslant&1,
\end{eqnarray}
and
\begin{eqnarray}\label{DC52}
	\mathcal{S}_{\mathsf{(b)}}=|I^{\mathsf{(b)}}_1(5,2)|^{\frac{1}{2}}+|I^{\mathsf{(b)}}_2(5,2)|^{\frac{1}{2}}&\leqslant&1,
\end{eqnarray} 
respectively. The terms are defined as $I^{\mathsf{(a)}}_j(N,l)=\Big\langle\prod_{u\in\{2,3\}}A_j^{u}\prod_{i\in\{1,4,5\}}\Delta_j^{i}\Big\rangle$ and $I^{\mathsf{(b)}}_j(N,l)=\Big\langle\prod_{u\in\{2,3,4\}}A_j^{u}\prod_{i\in\{1,5\}}\Delta_j^{i}\Big\rangle$ 
with $\Delta_j^{i}=\frac{1}{2}(A_2^{i}+(-1)^jA_1^{i})$, $j=1,2$.

Similar to the proof in  Sec.~\ref{appe31}, one can show that the inequality $\mathcal{S}_{\mathsf{(a)}}\leqslant\sqrt{2}$ can be violated when the inequality $\mathcal{S}_{\mathsf{(b)}}\leqslant\sqrt{2}$ holds. 
So two networks in Fig.~\ref{D1} can be distinguished by violating Eq.~(\ref{DQ53}) using quantum correlations achievable in network in Fig. \ref{D1}(b). 

\end{widetext}

\end{document}